\documentclass[pre,aps,twocolumn]{revtex4}
\usepackage{times}
\usepackage{amsmath}
\usepackage{graphicx}
\usepackage{amssymb}
\usepackage{color}
\usepackage{amsfonts}

\newcommand{\e}{\mathrm{e}}

\begin{document}

\title{Speed-of-light pulses in a nonlinear Weyl equation}
\author{Jes\'us Cuevas-Maraver$^{1}$}
\author{P. G. Kevrekidis$^{2}$}
\author{Franz G. Mertens$^{3}$}
\author{Avadh Saxena$^{4}$}
\affiliation{$^1$Grupo de Fisica No Lineal, Departamento de Fisica Aplicada I, Universidad de Sevilla, Escuela Polit\'ecnica Superior, C/ Virgen de \'Africa, 7, 40100-Sevilla, Spain \\
Instituto de Matem\'aticas de la Universidad de Sevilla (IMUS), Edificio Celestino Mutis, Avda. Reina Mercedes s/n, 41012-Sevilla, Spain }
\affiliation{$^2$Department of Mathematics and Statistics, University of Massachusetts, Amherst, MA 01003-4515, USA}
\affiliation{$^3$Physikalisches Institut, Universit\"at Bayreuth, D-95440 Bayreuth, Germany}
\affiliation{$^4$Center for Nonlinear Studies and Theoretical Division, Los Alamos National Laboratory, Los Alamos, New Mexico 87545, USA}
\begin{abstract}
  We introduce a prototypical
  nonlinear Weyl equation, motivated by recent developments
  in massless Dirac fermions, topological semimetals and photonics. We study the dynamics of its pulse solutions and find that a localized one-hump initial condition splits into a localized two-hump pulse, while an associated
  phase structure emerges in suitable components of the spinor field. For times larger than a transient time $t_s$ this pulse moves with the speed of light (or Fermi velocity in Weyl semimetals),
  effectively featuring linear wave dynamics and maintaining its shape (both in two and three dimensions). We show that
  for the considered nonlinearity,
  this pulse represents an exact solution of the nonlinear Weyl (NLW) equation.
  Finally, we comment on the generalization of the results to a broader
  class of nonlinearities and on their emerging potential for observation
  in different areas of application.
\end{abstract}
\maketitle

\section{Introduction}

The massless analogue of the Dirac equation, known as the Weyl equation, has seen a surge of interest in recent years due to a variety of newly discovered materials called Weyl semimetals \cite{RMP}, e.g. NbAs and TaP \cite{Weyl1, Weyl2}.  These harbor chiral quasiparticles called Weyl fermions and possess topological surface states~\cite{Weyl3}.  Weyl fermions exhibit linear dispersion, just like graphene, but are massless. In fact, Weyl semimetals are the three-dimensional (3D) analogues of graphene with broken spatial inversion or time reversal symmetry.  In the Brillouin zone of such materials, linear dispersion
arises around certain nodes, the so-called Weyl points, which always occur in pairs.  In addition, in Weyl semimetals regions described by different Chern numbers are connected by unclosed lines, the so-called Fermi arcs~\cite{Weyl4}, which can be experimentally observed using angle-resolved photoemission spectroscopy.  The Fermi arc starts from one Weyl point and ends at the other one with opposite chirality. The Weyl points (or nodes) are essentially monopoles of the quantized Berry flux in the crystal momentum or reciprocal space. Photonic counterparts of Weyl semimetals have been observed in double-gyroid structures using angle-resolved microwave transmission measurements~\cite{Weyl5}.   When time reversal and spatial inversion symmetries coexist in such a material, a pair of degenerate Weyl points may exist resulting in a Dirac semimetal~\cite{, RMP, Weyl4}, e.g. Cd$_3$As$_2$ \cite{Wang1} and Na$_3$Bi \cite{Wang2}.

In parallel to these developments of chiefly {\it linear} Weyl physics,
there has been an explosion of interest in the phenomenology of the
nonlinear version of the Dirac equation and its solitary waves;
a recent survey of the pertinent phenomenology can be found in~\cite{our}.
While the relevant model in its massive Thirring form~\cite{thirring}
was of interest to integrable systems and its Gross-Neveu/Soler
form~\cite{gross,soler} led to extensive studies in solitary waves
and their stability~\cite{our}, arguably, part of the recent appeal of the
model has been due to its applicability to a number of relevant physical setups.
Among these, we note the dynamical evolution of Bose-Einstein
condensates in the presence of honeycomb optical
lattices~\cite{Carr1,Carr2,Carr3,Carr4},
as well as the analogous propagation of light in honeycomb photorefractive
lattices, the so-called photonic graphene~\cite{peleg,ablo4,ablo3}.
These, in turn, motivated numerical and theoretical studies on the properties
of these models and revealed crucial differences from their nonlinear
Schr{\"o}dinger cousins, including, e.g., the potential absence of
the collapse instability for suitable parametric intervals in two-dimensional
systems~\cite{PRL}.

In light of these developments, it is natural to consider
a prototypical model that would be suitable for the analogous
Weyl systems (most notably so in 3 spatial dimensions), especially
given that some of the considered systems, such as the optical
ones of~\cite{Weyl5} are settings where the tuning of optical intensity
may lead to the controllable introduction of nonlinearity. It is
worthwhile to also note a recent motivation of the notion of
nonlinear Weyl media at the discrete level (as opposed to the
prototypical continuum formulation herein) in the context
of atomic Bose-Einstein condensates in the work of~\cite{borism}.
It is the introduction of such a model blending the underlying
linear Weyl operator (i.e., a massless 3D Dirac operator --although
we also consider the 2D analogue thereof as well--) and
a cubic nonlinearity that we explore in the present setting.
Given the extensive number of corresponding studies at the Dirac
level,
but also its properties under Lorentz transformations and
remarkable phenomenology reported below,
we select the Gross-Neveu/Soler type of nonlinearity to formulate
  a  nonlinear Weyl equation model that may be a
  starting point for exploring
  the interplay of nonlinearity with linear Weyl
  operators in various contexts. The analytical
  (and numerical) results obtained herein can operate as a guide
  for examining other nonlinearities including the more relevant
for atomic condensates cases of a Kerr type nonlinearity~\cite{borism}.

Our presentation and main results are as follows. First,
we formulate the 3D nonlinear Weyl (NLW) equation and present some of
its principal properties in 3 spatial dimensions, including
most notably the observation that pulse-like initial data split
into a two-humped ring density structure (acquiring a suitable phase in
some of the spinor components). Beyond a transient time, the
resulting density excitation is found to propagate at the speed of
light (or Fermi velocity in Weyl semimetals). We demonstrate that, as a consequence, the resulting waveforms
satisfy an effective 3D wave equation which is analytically solvable
via suitable transformations. To corroborate these findings,
we also examine the corresponding 2D case, and demonstrate
the generic nature of the relevant phenomenology.
It is important to point out here that our results bear fundamental
  differences from the recent 1D corresponding study of~\cite{NiurkaPRE}.
  In particular, here (a) there is a single radial density structure (as opposed
  to two pulses in the 1D case); (b) there appears a phase (vorticity)
  profile that we discuss below and finally (c) the density decays with
  the distance, features that are particular to the higher dimensional
  settings.
Upon elucidating these traits, we
summarize our main results, offer a number of remarks
regarding other nonlinearities and applications, and propose a
number of associated directions for future study.

\section{NLW Model}

One of our principal motivations for utilizing the Gross-Neveu/Soler
nonlinearity is that the associated quantity $\overline{\psi}\psi$
transforms as a scalar under the Lorentz transformation.
This naturally suggests the corresponding massless Lagrangian density:
\begin{equation}\label{nlw-lagrangian}
\mathcal{L}\sb{\mathrm{Weyl}}
=
\bar\psi \left(i\gamma\sp\mu\partial\sb\mu \right) \psi + F\left(\bar\psi\psi\right),
\end{equation}
where $\psi(x,t)\in\mathbb{C}^N$,
$x\in\mathbb{R}^n$
and $\gamma\sp\mu$,
$0\le\mu\le n$,
are $N\times N$ Dirac $\gamma$-matrices
satisfying the anticommutation relations
$\{\gamma\sp\mu,\gamma\sp\nu\}=2\eta\sp{\mu\nu}$,
with $\eta\sp{\mu\nu}$ the Minkowski tensor~\cite{dewit},
and $\bar\psi=\psi\sp\dagger\gamma\sp 0$.

Since the Weyl physics emerges in 3D settings, it is natural
to start by considering the three dimensional case in which the spinors have four components. The NLW equation derived from the Lagrangian density
of Eq.~(\ref{nlw-lagrangian}), in Cartesian coordinates assumes the form:
\begin{eqnarray}\label{eq:NLW3D}
    i\partial_t\psi_1 &=& -i[(\partial_x-i\partial_y)\psi_4+\partial_z\psi_3]-f(\bar\psi\psi)\psi_1 ,\nonumber \\
    i\partial_t\psi_2 &=& -i[(\partial_x+i\partial_y)\psi_3-\partial_z\psi_4]-f(\bar\psi\psi)\psi_2 ,\nonumber \\
    i\partial_t\psi_3 &=& -i[(\partial_x-i\partial_y)\psi_2+\partial_z\psi_1]+f(\bar\psi\psi)\psi_3 ,\nonumber \\
    i\partial_t\psi_4 &=& -i[(\partial_x+i\partial_y)\psi_1-\partial_z\psi_2]+f(\bar\psi\psi)\psi_4 ,
\end{eqnarray}
with $F'\left(\bar\psi\psi\right)=f(\bar\psi\psi)$ which here is chosen as $f(\bar\psi\psi)=g(|\psi_1|^2+|\psi_2|^2-|\psi_3|^2-|\psi_4|^2)$;
we use the value of the prefactor $g=1$ herein. Notice that, contrary to the linear case, the transformation of the 4-spinor $\psi$ into Weyl 2-spinors with left and right chirality does not decouple the equations.

We have employed the Wakano ansatz~\cite{Wakano}
\begin{equation}
\label{eq:ansatz3D}
    \psi(\vec r,0)=\phi(\vec r)=\left[\begin{array}{c}
    u(r) \\
    0 \\
    i\,v(r)\cos\theta \\
    i\,v(r)\sin\theta\mathrm{e}^{i\varphi}
    \end{array}\right]
\end{equation}
\noindent to initialize the NLW equation and the spherical frame will be useful in our analytical considerations below. Nevertheless, for our numerical solution of Eq.~(2), we use the Fourier spectral collocation method in Cartesian coordinates adapting the method used in~\cite{PRL} to the 4-spinor case in 3D (see Appendix~\ref{sec:numerical}). The total mass stemming from the integration over space of the mass density
\begin{equation}
  \rho(r,t)=|\psi_1(r,t)|^2+|\psi_2(r,t)|^2+|\psi_3(r,t)|^2+|\psi_4(r,t)|^2
  \label{dens}
\end{equation}
is a conserved quantity of the model.

\begin{figure}
\begin{center}
\includegraphics[width=8cm]{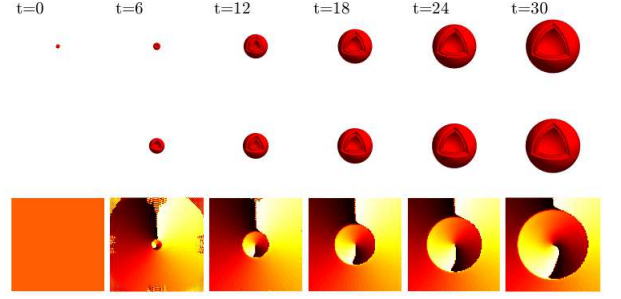}
\end{center}
\caption{Snapshots showing the evolution of an initial hump in the 3D NLW equation. Top (middle) row shows an isosurface for $0.25$ times the maximum of
  $|\psi_1(\vec r,t)|^2+|\psi_2(\vec r,t)|^2$ ($|\psi_3(\vec r,t)|^2+|\psi_4(\vec r,t)|^2$) at different values of time $t$; {an octant of the sphere has been removed in order to get a better visualization of the two-humped nature of the
  resulting structure}. The bottom row shows the phase of $\psi_4(\vec r,t)$ with emerging vorticity. {In each picture, the axes cover the range $[-40,40]$}.}
\label{fig:collage3D}
\end{figure}

\begin{figure}
\begin{center}
\includegraphics[width=8cm]{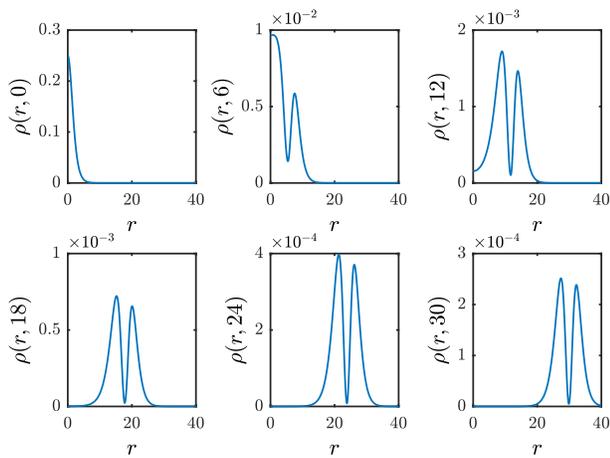}
\end{center}
\caption{Density $\rho(r,t)$ at different values of time in the 3D NLW equation.}
\label{fig:NLW3D1}
\end{figure}

\begin{figure}
\begin{center}
\begin{tabular}{cc}
\includegraphics[width=4cm]{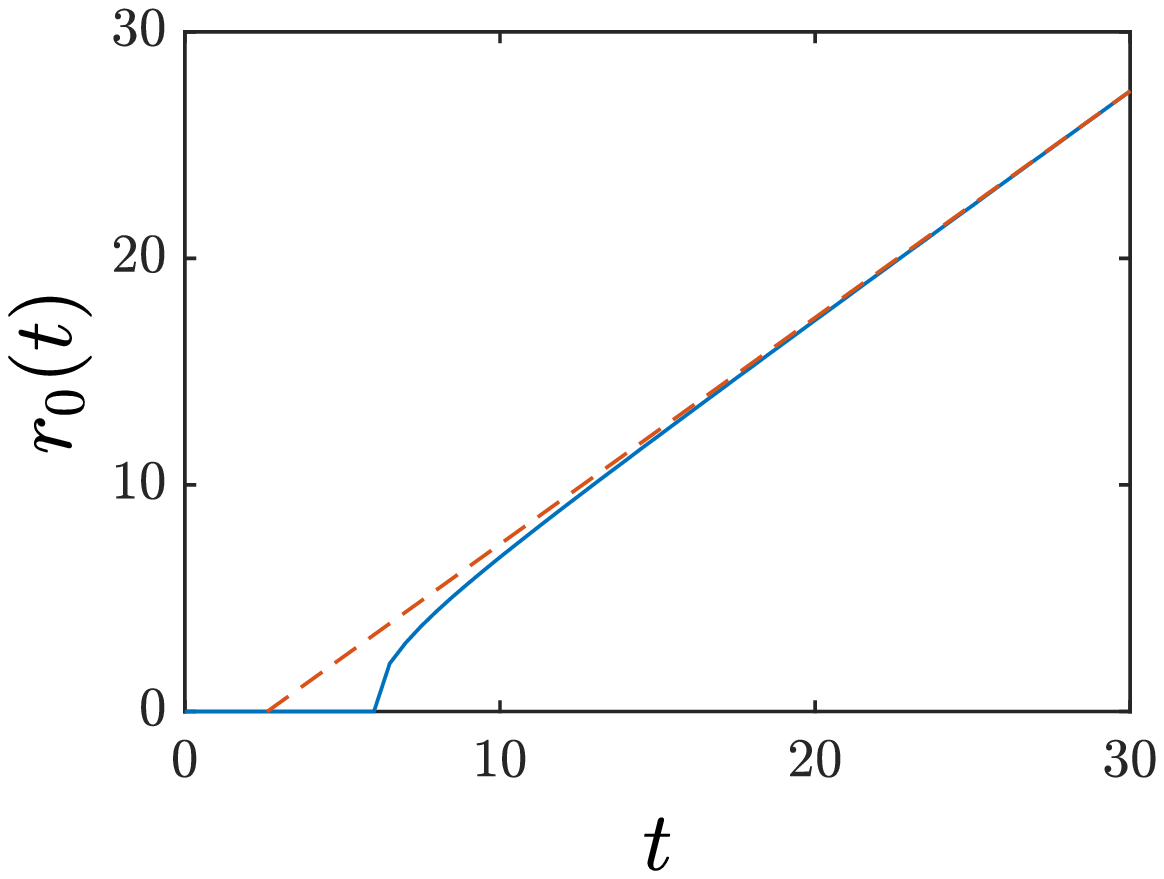} &
\includegraphics[width=4cm]{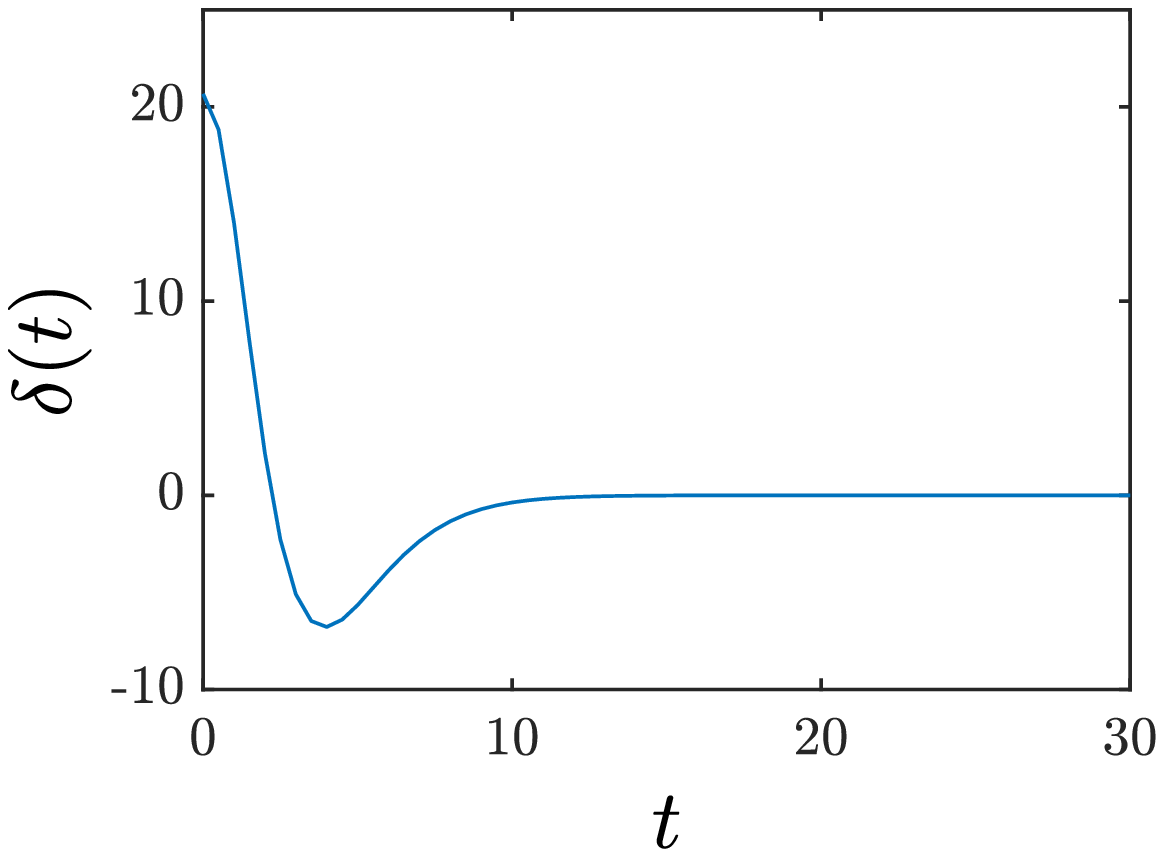}
\end{tabular}
\end{center}
\caption{The left panel shows the position of the {leftmost} density {local} maximum $r_0$ for the 3D NLW equation; the dashed red line corresponds to a slope 1 line to which $r_0(t)$ tends asymptotically. The right panel shows the evolution of $\delta(t)$ [see Eq. (\ref{eq:delta})], showcasing its asymptotic vanishing. {Notice that we have taken $r_0=0$ when there is a single local maximum.}}
\label{fig:NLW3D2}
\end{figure}

We now integrate the NLW model of Eq.~(\ref{eq:NLW3D})
for typical pulse-like initial data of the form:
\begin{equation}\label{eq:initial}
    u(r)=\frac{1}{2}\mathrm{sech}\frac{r}{2}, \qquad v(r)=0.
\end{equation}
Notice that we have found similar results for other forms
of such initial data (e.g. Gaussian, etc.), and also by taking nonzero $v(r)$.
Figure~\ref{fig:collage3D} shows snapshots of
(isocontour density, as well as phase profiles of) both $|\psi_1(\vec r,t)|^2+|\psi_2(\vec r,t)|^2$ and $|\psi_3(\vec r,t)|^2+|\psi_4(\vec r,t)|^2$, i.e., the
densities of two component pairs. Note that the initial localized hump at the first spinor component transforms into a spherical shell two-hump structure
that expands with time, whereas the initially null third and fourth spinor components also transform into a similar pattern, with the latter displaying vorticity, in line with
the Wakano ansatz of Eq.~(\ref{eq:ansatz3D}); the second spinor component
remains null (within machine precision) during the dynamical evolution.
Figure~\ref{fig:NLW3D1} shows the density
at different times; notice the persistence of the two local maxima
over the propagation time.
The left panel of Fig.~\ref{fig:NLW3D2} shows the position of the local
density maximum; from this it is evident that the `ring' expands asymptotically with speed 1 beyond a transient time, i.e., for $t>t_s$.

The right panel of Fig.~\ref{fig:NLW3D2} shows the quantity
\begin{equation}\label{eq:delta}
    \delta(t)=\int \mathrm{d}^N\vec{r} f(\psi(\vec r,t)),
\end{equation}
which tends to zero for $t>t_s$, with $N$ being the number of spatial dimensions of the system.
This quantity is a measure of the nonlinearity of the system during
time evolution.
It is thus clear that the nonlinear term becomes effectively ``deactivated''
for $t>t_s$ (in line also with the 1D massless Dirac case findings
of~\cite{NiurkaPRE}). As a result, the emerging two-humped pulses
propagate at the ``speed of light'' or ``Fermi velocity'', effectively satisfying the
{\it linear} 3D wave equation, given the spontaneous vanishing of the
nonlinear term $f$. In other words, given our observation that $f\rightarrow0$, it is straightforward
to show that each spinor $U=\psi_{1,2,3,4}$ satisfies the linear
3D wave equation of the form:
\begin{equation}
\left(\frac{1}{c^2}\partial_t^2 - \partial_x^2 - \partial_y^2 - \partial_z^2 \right) U = 0 ,
\end{equation}
for which the transformation $w = rU$ can factor out the curvature
term $(2/r)\partial_rU$ and effectively restore a 1D wave equation in the radial
variable, ultimately retrieving the full solution in the form:
\begin{equation}
U(r,t) = \frac{1}{r} [h^{(1)}(r-ct) + h^{(2)}(r+ct)] .
\end{equation}
In our simulations $h^{(2)}=0$ and for $t\ge t_s$ the four spinor components $\psi_i=U_i=\frac{1}{r} h_i^{(1)}(x-ct)$ with four functions ($i=1,\dots,4$)
produce the two-hump structure seen in the density, per Eq.~(\ref{dens}), in Fig.~\ref{fig:NLW3D1}.

Remarkably, this two-hump structure is a unique, previously undiscovered feature which differs qualitatively from the two-hump structure that was observed in the 1D NLW equation \cite{NiurkaPRE}. Here the initial pulse splits symmetrically into two equal humps which move in opposite directions with the speed of light (or Fermi velocity).
From a materials perspective, another unique 3D feature is that TaAs has 12 pairs of Weyl nodes: four pairs lie in the $k_z=0$ plane in the Brillouin zone (above the Fermi energy) and the remaining eight pairs are located off the $k_z=0$ plane (below the Fermi energy). An interesting aspect of our results would be to explore which nodes and Fermi arcs are affected during the propagation of the pulse and its subsequent splitting.

\begin{figure}
\begin{center}
\includegraphics[width=8cm]{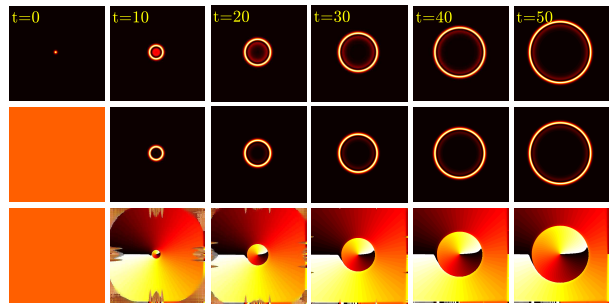}
\end{center}
\caption{Snapshots showing the evolution of an initial hump in the 2D NLW equation. Top (middle) row shows the value of $|\psi_1(\vec r,t)|^2$ ($|\psi_2(\vec r,t)|^2$) at different values of time $t$. Bottom row shows the development of a vortical phase structure within $\psi_2(\vec r,t)$. {In each picture, the axes cover the range $[-80,80]$}.}
\label{fig:collage2D}
\end{figure}

\section{2D NLW equation}

For the sake of comparison with the 3D case and a better understanding of NLW in general, next we study the 2D NLW equation, which also
showcases the generality of our findings. In this context, we note that based on the non-symmorphic space group symmetries 2D Dirac semimetals have been predicted \cite{2DSM1} and SrMnSb$_2$ has been proposed as a candidate material \cite{2DSM2}.  However, 2D Weyl semimetals are not supposed to exist in nature due to the absence of ``chiral anomaly" in 2D; moreover the Weyl nodes are unstable in 2D. Nonetheless, there is a recent proposal that a HgTe/CdTe quantum well structure may exhibit the 2D Weyl semimetallic behavior for certain thickness of HgTe layers sandwiched between a normal insulator and a quantum spin Hall insulator due to a twofold rotation symmetry about the growth direction \cite{2DWSM}.

In 2D the $\gamma$-matrices are defined as $\gamma^0 = \sigma_3$ and $\gamma^j = \sigma_3 \sigma_j$ with $j=1,2$, where $\sigma_1$, $\sigma_2$ and $\sigma_3$ are the Pauli matrices. Explicitly, $\gamma^1 = \sigma_3\sigma_1 = i\sigma_2$
and $\gamma^2 = \sigma_3\sigma_2$.
In this context, the simplest case example of interest
derived from the Lagrangian density in Eq.~(\ref{nlw-lagrangian})
can involve solely two spinor components according to the dynamical equations \cite{PRL}:
\begin{eqnarray}\label{eq:dyn}
    i\partial_t\psi_1 &=& -(i\partial_x+\partial_y)\psi_2 - f(\bar\psi\psi)\psi_1 ,\nonumber \\
    i\partial_t\psi_2 &=& -(i\partial_x-\partial_y)\psi_1 + f(\bar\psi\psi)\psi_2 ,
\end{eqnarray}
where $\psi_1$, $\psi_2$
are the components of the spinor $\psi\in\mathbb{C}^2$ and
the nonlinearity is
$f = g(|\psi_1|^2 - |\psi_2|^2)$. We note that Eq.~(\ref{eq:dyn}) is a $\mathbf{U}(1)$, as well as
translation-invariant, Hamiltonian system.

In Fig.~\ref{fig:collage2D}, we have once again explored the evolutionary
dynamics of the 2D analogue of the NLW equation, initializing with a
single humped waveform.
The relevant results are, once again, generic
in their nature within the class of such initial data. We observe here too
that a two-humped structure spontaneously emerges in a ``ring'' form
(for the density),
with the 2nd component also featuring a phase profile, associated with
the presence of vorticity in this spinor component.
Similar to the 3D case, and showcasing the
generality of our observations, we find that for $t\ge t_s \simeq 20$,
the pulses propagate with constant speed, namely the speed of light or Fermi velocity,
and the nonlinearity once again is made to vanish due to
$|\psi_1|^2=|\psi_2|^2$, leading to an effectively {\it linear} dynamics.

In a calculation similar to the above 3D case, given
in detail in Appendix~\ref{sec:2D} (where both standing and traveling
wave solutions of the 2D NLW are explored, as applicable in the
case of $t\ge t_s$), we find that the effective dynamics for $t\ge t_s$
amounts to:
\begin{eqnarray}
  \partial_t \psi_1 + (\partial_x - i\partial_y)  \psi_2 &=& 0 ,
  \label{2dnlw1}
  \\
\partial_t \psi_2 + (\partial_x + i\partial_y) \psi_1 &=& 0.
\label{2dnlw2}
\end{eqnarray}
Combining the two equations (by taking, e.g., a time-derivative of
the first and substituting in the second), we obtain a 2D wave equation for both
$\psi_1$ and $\psi_2$, which, in turn, leads to the following expression for
the density:
\begin{equation}
\rho(r,t) =  \frac{1}{r} |f(r-t)|^2.
\end{equation}
Constant factors are omitted here because $\psi_1$ and $\psi_2$ are solutions
of effective linear equations. Notice the important $1/r$ effect, induced
by the presence of the curvature also in the 2D system; such a term
would be absent in a massless 1D Dirac setting~\cite{NiurkaPRE}.

\section{Conclusions, Extensions and Future Work}

Motivated by the recent discovery of Weyl semimetals in NbAs and TaP \cite{RMP, Weyl1, Weyl2} and their photonic analogue~\cite{Weyl5} we have introduced a prototypical nonlinear Weyl (NLW) equation in 3D (and examined
its analogue in 2D), which is the {\it massless} variant of the nonlinear Dirac equation (NLD) for the 3D case. Given its invariance under Lorentz
transformations, we have utilized the Gross-Neveu/Soler nonlinearity in this study.
We have obtained pulse solutions of NLW and their time evolution.
Beyond a transient time $t>t_s$, we have found that
these pulses move with the speed of light and satisfy an effectively
linear (and explicitly solvable) wave equation. In real Weyl semimetals the speed of light should be replaced by the Fermi velocity $v_F$ \cite{RMP}.
In the process,
the role of curvature in the evolution of these pulses, as well
as their two-humped structure and spontaneous phase development in
suitable components, have also been elucidated. The results on 2D NLW equation
may, in principle, be relevant to the possibility of a 2D Weyl semimetal in HgTe/CdTe quantum well structures \cite{2DWSM}.

We recognize that the model presented here has not been firmly tied to the phenomenology
of NLW fermions. However, we believe it is of interest in its own right as a prototypical equation
that would bear the main ingredients of nonlinearity and Weyl physics (i.e., massless 3D, but
also additionally 2D). Thus, our findings are likely to be of interest to multiple areas of
physics.  In this regard, an experimental study of the optical nonlinear response and measurement
of optical conductivity of the Weyl semimetal TaAs shows a saturable characteristic
at very large intensity~\cite{ChiAP}.  The optical field can also enhance the
fermion mobility because of the interaction of Fermi arcs and Weyl nodes under the
influence of the optical field. This observation points to the importance of Weyl
fermion dynamics in photolectronics and optoelectronics which,
in turn, are among the areas motivating the present study.
Specifically,
our results may bear connections with
how the dynamics of the pulse (e.g., initialized at the surface)
and its splitting affect the
topological surface states in a Weyl semimetal.
Similarly, collisional nonlinearity of atomic
Bose-Einstein condensates in ultracold atomic gases \cite{borism} also provides
a possible realization of NLW fermions and the pulse solutions studied here.

Our results provide insight
into the localization and dynamics of {\it massless} Dirac fields in the presence
of nonlinearity. However, they also pose important questions that are
especially relevant to address in future studies. In particular,
from the theoretical standpoint, while the Gross-Neveu/Soler nonlinearity
is of interest given its symmetry properties, in optical and atomic
Dirac settings a nonlinearity involving solely $|\psi_i|^2 \psi_i$
in the equation for the $i$-th spinor (i.e., a Kerr effect solely
in each component from its own self-action) is naturally of interest.
It is then particularly relevant to separately explore the latter
situation in both 3D and 2D.
Remarkably our preliminary observations suggest that in these settings
too, {\it despite} the variation of the nonlinearity, a similar phenomenology
is observed. Namely, long-lived pulses in the form of envelope solitons
appear to propagate outward at the speed of light or Fermi velocity for the class of
initial data considered herein.
Moreover, it would be particularly interesting
in experimentally realized double-gyroid photonic crystals with broken inversion
symmetry~\cite{Weyl5} or perhaps in theoretically proposed atomic settings to
explore the possibility of observing this intriguing interplay of linear Weyl
phenomena (some of which have been discussed above) and nonlinearity.
Finally, it would be insightful to compare solitons in Dirac and Weyl fermion
systems with those related to the {\it third} kind of fermion, namely Majorana \cite{Majorana}.

\begin{acknowledgements}
J.C.-M. thanks financial support from MAT2016-79866-R project (AEI/FEDER, UE). P.G.K. acknowledges support from NSF-PHY-1602994.
F.G.M. acknowledges the hospitality of the Center for Nonlinear Studies and Theoretical Division at LANL.  This work was supported in part by the US Department of Energy. We acknowledge the useful comments of Andrew Comech.
\end{acknowledgements}

\appendix

\section{Numerical methods}
\label{sec:numerical}

We briefly describe in this appendix the numerical methods employed for integrating equations (\ref{eq:NLW3D}). For a more complete description of such methods, the reader is referred to \cite{our}.

The first step to follow is to implement a grid and a method for discretizing the spatial derivatives of the partial differential equations (PDEs) of (\ref{eq:NLW3D}) and, consequently, transforming them into a set of coupled ordinary differential equations. Finite difference methods do not usually work well in nonlinear Dirac equations. Instead, one must make use of spectral methods. In our case, as we are dealing with numerical integrations of PDEs and the pulse tends to infinity in an exponential way, a well-suited choice is the Fourier spectral method. Such a method requires the use of periodic boundary conditions and an equispaced grid. The implementation is quite simple, as it basically consists of performing direct and inverse Fourier transforms. That is, if we denote by $U(x,y,z)$ any of the spinor components $\psi_{1,2,3,4}(x,y,z)$ then the derivative with respect to, e.g. direction, $x$  is given by
\begin{equation*}
    \partial_x U(x,y,z)=\mathcal{F}^{-1}_x\left(ik\mathcal{F}_x(U(x,y,z))\right) \,,
\end{equation*}
where $\mathcal{F}_x$ and $\mathcal{F}^{-1}_x$ denote the direct and inverse, respectively, one-dimensional Fourier transform in the direction $x$. Notice that $U(x,y,z)$ actually represents an $N\times N\times N$ array which only takes values at the grid points. Because of this, Fourier transforms can be accomplished by means of the Fast Fourier Transform (FFT) and $k\equiv\{k_n\}$ is a vector with $N$ components given by $k_n=n\pi/L$ for $n<N$ and $k_N=0$.

Once we have defined our set of ordinary differential equations, the second ingredient is the numerical integrator. In our case, the integrator we prefer to use for simplicity and accuracy is the Dormand-Prince~\cite{hairer} algorithm.

All the above schemes have been implemented using Matlab in a desktop PC with 8 Gb of RAM. In our particular case, the domain has been a box of size $(-L,L]\times(-L,L]\times(-L,L]$ with $L=48$ and the lattice discretization parameter is $h=0.8$; with these data, the number of grid points is $N=120$ so that the total number of ordinary differential equations to integrate is equal to $4N^3\sim7\times10^6$. Reaching the value of time $t=30$ takes around a day. We finally remark that one of the advantages of using spectral methods is that it works even with a rough discretization and that the value of the wavefunction at different values of $(x,y,z)$ out of the grid can be accurately attained by using spline interpolation.

\section{The two-dimensional massless Nonlinear Dirac case}
\label{sec:2D}

For the 2D case, we can simplify the relevant analysis
by using the polar coordinates, where the equations take the form
\begin{eqnarray}\label{eq:stat}
    i\partial_t\psi_1
&=& -\e^{-i\theta}\left(i\partial_r+\frac{\partial_\theta}{r}\right)\psi_2 - f(\psi_1,\psi_2)\psi_1,
\nonumber \\
    i\partial_t\psi_2
&=& -\e^{i\theta}\left(i\partial_r-\frac{\partial_\theta}{r}\right)\psi_1 + f(\psi_1,\psi_2)\psi_2.
\end{eqnarray}
We have performed simulations of the relevant 2D analogue of the
NLW model with the following initial conditions, in line with those
used in Ref.~\cite{PRL} (for the massive case),
\begin{equation}
\label{eq:ansatz2D}
    \psi(\vec r,0)=\phi(\vec r)=\left[\begin{array}{c}
    u(r)\e^{i S\theta} \\
    i\,v(r)\e^{i(S+1)\theta}
    \end{array}\right]
\end{equation}
with $S$ being the vorticity and, just as in the 3D case,
\begin{equation}\label{eq:initial2d}
    u(r)=\frac{1}{2}\mathrm{sech}\frac{r}{2}, \qquad v(r)=0 .
\end{equation}

\begin{figure}[ht]
\begin{center}
\includegraphics[width=8cm]{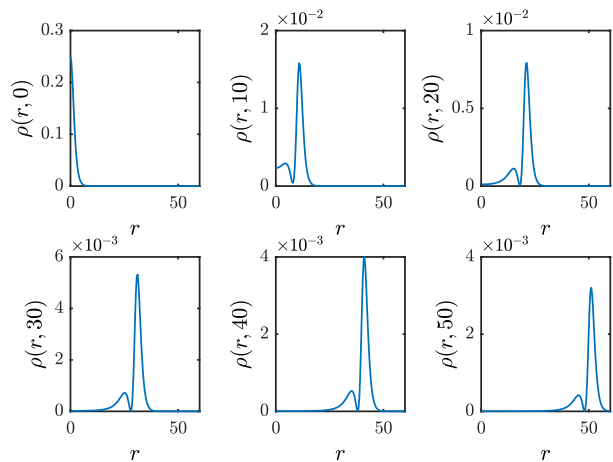}
\end{center}
\caption{Density $\rho(r,t)$ at different values of time in the 2D massless nonlinear Dirac equation.}
\label{fig:NLW2D1}
\end{figure}

\begin{figure}[ht]
\begin{center}
\begin{tabular}{cc}
\includegraphics[width=4cm]{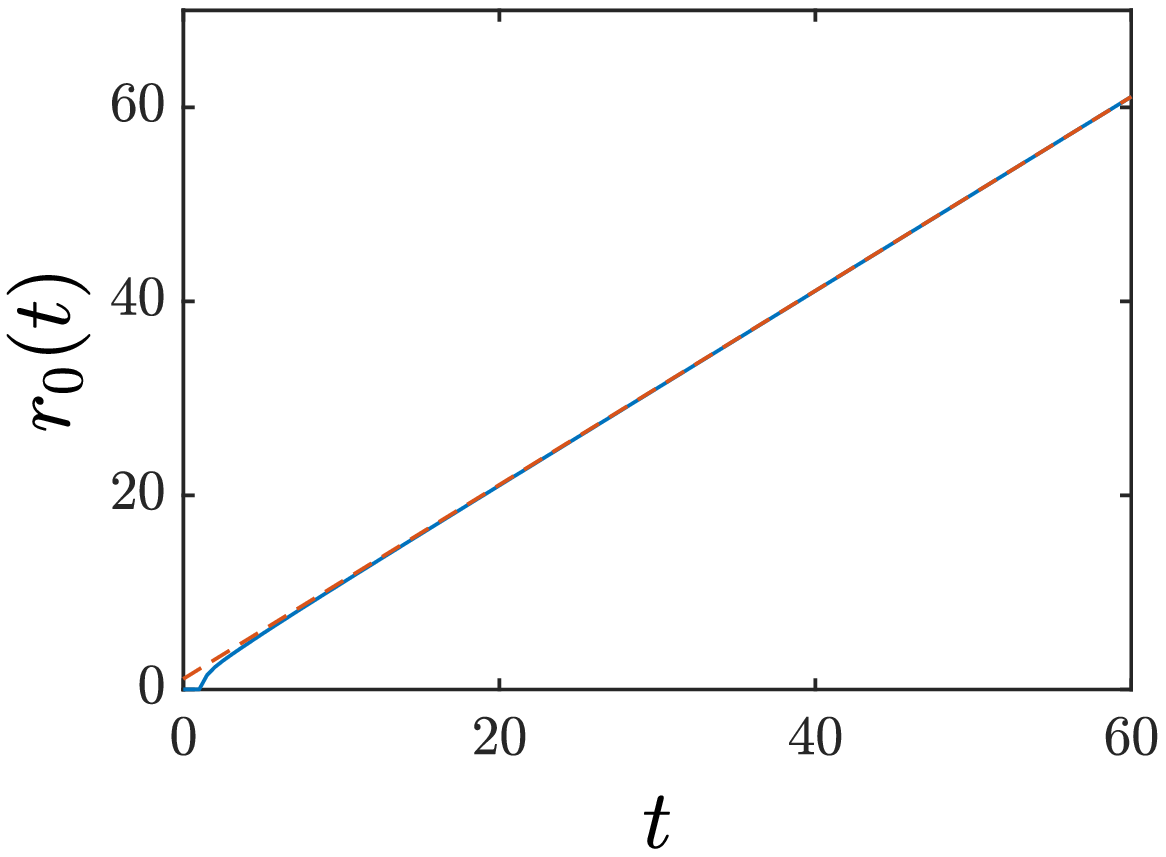} &
\includegraphics[width=4cm]{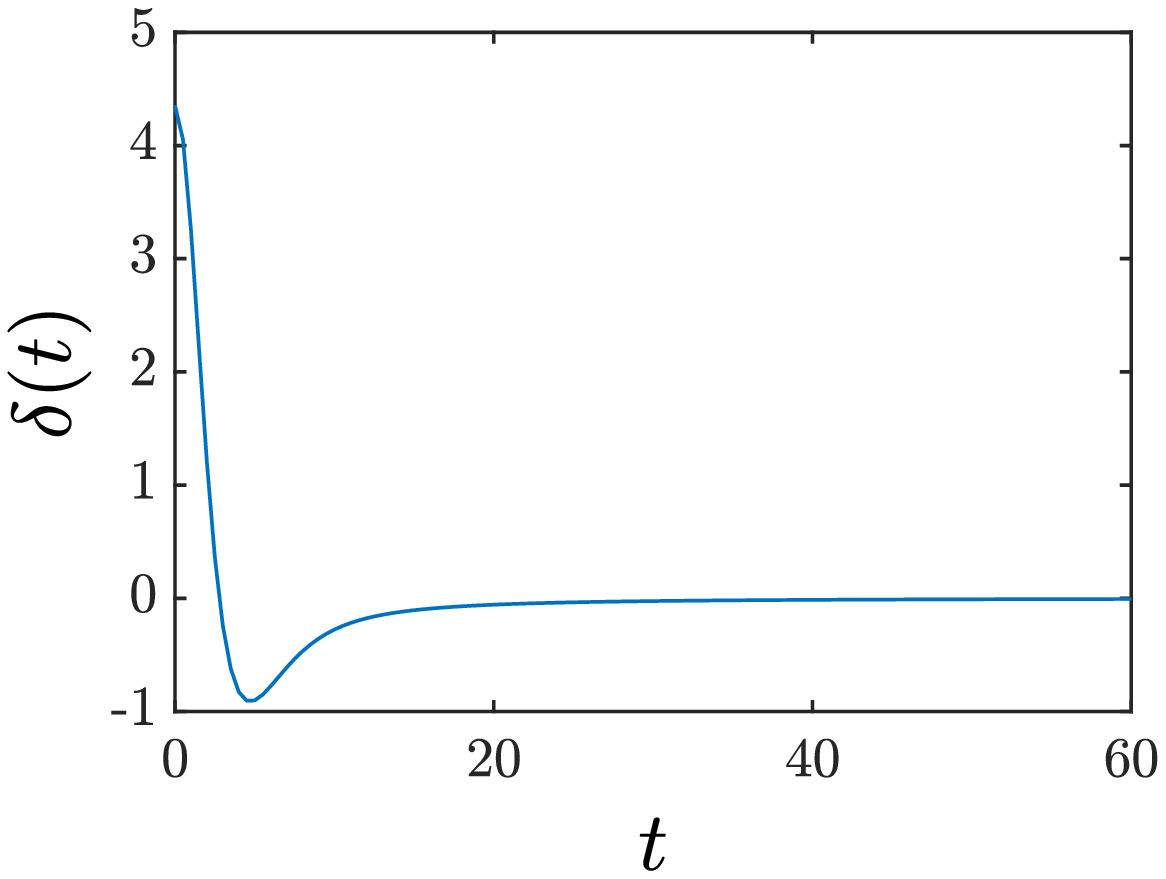}
\end{tabular}
\end{center}
\caption{Left panel shows the position of the density maximum $r_0$ for the massless 2D nonlinear Dirac equation; dashed red line corresponds to a slope 1 line to which $r_0(t)$ tends asymptotically. Right panel shows the evolution of $\delta(t)$ and its effective vanishing beyond a transient time.}
\label{fig:NLW2D2}
\end{figure}

The two-humped ring nature of the resulting dynamics (as well
as the spontaneous emergence of vorticity) is revealed
in the figure shown in the main text. Here, for completeness we show
in Fig.~\ref{fig:NLW2D1} the density $\rho(r,t)$ at different times.
Also, the left panel of Fig.~\ref{fig:NLW2D2} shows the position of the
maximum of the density of the ring; from this figure, it is evident that
the ring expands asymptotically with the speed of light or Fermi velocity
(as in 3D). The right panel of
the figure shows $\delta(t)$ defined in the main text. The vanishing of
this quantity once again indicates the spontaneous ``self-annihilation''
of the nonlinear terms. Thus, for $t>t_s$ this structure remains the same for all times, but its
density is reduced by an $r$-dependent factor and it moves with the speed of light or Fermi velocity.

Therefore, this structure is supposed to be a solution of the linear
2D form of the equation. In the following we will show that it is indeed the case by solving exactly the associated linear PDEs.
We use the following Ans\"atze
\begin{equation}
\psi_1 = e^{-i\frac{\theta}{2}} \tilde{\psi_1}(r,t) ,  ~~~ \psi_2 = e^{i\frac{\theta}{2}} \tilde{\psi}_2(r,t) ,
\end{equation}
and obtain the following relations:
\begin{eqnarray}\label{eq:1}
\partial_t \tilde{\psi_1} = - \left(\partial_r + \frac{1}{2r}\right) \tilde\psi_2 ,  \nonumber \\
\partial_t \tilde{\psi_2} = - \left(\partial_r + \frac{1}{2r}\right) \tilde\psi_1 .
\end{eqnarray}

First we consider time independent solutions:
\begin{equation}
\tilde{\psi_1}(r,t) = u(r), ~~~ \tilde{\psi_2}(r,t) = iv(r) ,
\end{equation}
and obtain
\begin{equation}
v' + \frac{1}{2r} v =0 ,  ~~~ u' + \frac{1}{2r} u = 0 .
\end{equation}
We find that
\begin{equation}
v = c_2 e^{-\int \frac{1}{2r} dr} = c_2 e^{-\frac{1}{2}\ln r} = \frac{c_2}{\sqrt{r}} ,
\end{equation}
with arbitrary $c_2$.  In a similar way we have $u = c_1/\sqrt{r}$ with $c_1$ being an arbitrary constant. Thus,
\begin{equation}
\Psi = (\psi_1,  ~~\psi_2)^T = \frac{1}{\sqrt{r}} (c_1 e^{-i\frac{\theta}{2}},  ~~ ic_2 e^{i\frac{\theta}{2}} )^T.
\end{equation}

Denoting $\tilde u = \tilde \psi_1 + \tilde \psi_2$ and $\tilde v = \tilde \psi_1 - \tilde\psi_2$ we decouple Eqs. (\ref{eq:1}) as
\begin{eqnarray}\label{eq:2}
\partial_t \tilde u =  - \left(\partial_r + \frac{1}{2r}\right) \tilde u , \nonumber \\
\partial_t \tilde v =  +\left(\partial_r + \frac{1}{2r}\right) \tilde v .
\end{eqnarray}
Next, we consider stationary solutions by employing the ansatz
\begin{equation}
\tilde{u}(r,t) = e^{-i\omega t} u(r) , ~~~ \tilde{v}(r,t) = e^{-i\omega t} v(r)
\end{equation}
and obtain
\begin{equation}
u(r) = c_1 \frac{e^{i\omega r}}{\sqrt{r}} , ~~~ v(r) = c_2 \frac{e^{-i\omega r}}{\sqrt{r}} .
\end{equation}
Finally, we get the solutions
\begin{eqnarray}
\psi_1 =\frac{1}{2} e^{-i\omega t} e^{-i\theta/2} [u(r) + v(r)] , \nonumber \\
\psi_2 =\frac{1}{2} e^{-i\omega t} e^{+i\theta/2} [u(r) - v(r)] .
\end{eqnarray}

Next we consider traveling wave solutions.  We take the decoupled Eqs.~(\ref{eq:2}) and
make the ansatz
\begin{equation}
\tilde u = \frac{1}{r^{\beta}} h^{(1)}(r -ct) ,
\end{equation}
with a solution $c=1$, $\beta=1/2$, where $h^{(1)}(r-ct)$ is an arbitrary function, and
\begin{equation}
\tilde v = \frac{1}{r^{\beta}} h^{(2)}(r -ct) ,
\end{equation}
with a solution $c=-1$, $\beta=1/2$. That is, we have two solutions (with $h^{(1)}$ and $h^{(2)}$ arbitrary):
\begin{eqnarray}
\tilde{u}_I = \frac{1}{\sqrt{r}} h^{(1)}(r-ct) , ~~ c=1 , ~~~ \tilde{v}_I = 0 ,  \nonumber\\
\tilde{u}_{II} = 0, ~~~  \tilde{v}_{II} = \frac{1}{\sqrt{r}} h^{(2)}(r -ct) , ~~ c=-1 .
\end{eqnarray}
Thus, the solutions of the original equations (with $h^{(1)}$ and $h^{(2)}$ arbitrary)  are

\begin{equation}
\begin{split}
& \psi_{1,I} = \frac{1}{2} e^{{-i\theta}/2} \frac{1}{\sqrt{r}} h^{(1)}(r-t), \ \
\psi_{2,I} = \frac{1}{2} e^{{+i\theta}/2} \frac{1}{\sqrt{r}} h^{(1)}(r-t) . \\
& \psi_{1,II} = \frac{1}{2} e^{{-i\theta}/2} \frac{1}{\sqrt{r}} h^{(2)}(r+t),  \ \
\psi_{2,II} = \frac{1}{2} e^{{+i\theta}/2} \frac{1}{\sqrt{r}} h^{(2)}(r+t) .
\end{split}
\end{equation}


\begin{thebibliography}{99}

\bibitem{RMP} N. P. Armitage, E. J. Mele, and A. Vishwanath, Rev. Mod. Phys. {\bf 90}, 015001 (2018).

\bibitem{Weyl1} S.-Y. Xu et al., Science {\bf 349}, 613 (2015).

\bibitem{Weyl2} B. Q. Lv et al., Phys. Rev. X {\bf 5}, 031013 (2015).

\bibitem{Weyl3} A. A. Burkov and L. Balents, Phys. Rev. Lett. {\bf 107}, 127205 (2011).

\bibitem{Weyl4} O. Vafek and A. Vishwanath, Ann. Rev. Cond. Mat. Phys. {\bf 5}, 83 (2014).

\bibitem{Weyl5} L. Lu, Z. Wang, D. Ye, L. Ran, J. D. Joannopoulos, and M. Soljacic, Science {\bf 349}, 622 (2015).

\bibitem{Wang1} Z. Wang, H. Weng, Q. Wu, X. Dai, and Z. Fang, Phys. Rev. B {\bf 88}, 125427 (2013).

\bibitem{Wang2} Z. Wang, Y. Sun, X.-Q. Chen, C. Franchini, G. Xu, H. Weng, X. Dai, and Z. Fang, Phys. Rev. B {\bf 85}, 195320 (2012).

\bibitem{our} J. Cuevas-Maraver, N. Boussaid, A. Comech, R. Lan, P.G. Kevrekidis, A. Saxena, Springer Series in Understanding
Complex Systems (Eds. V. Carmona et al.), Nonlinear Systems v. 1, p. 89 (2018); arXiv:1707.01946.

\bibitem{thirring} W.E. Thirring,
  Ann. Phys. {\bf 3}, 91 (1958).

\bibitem{gross} D.J. Gross, A. Neveu,
  Phys. Rev. D {\bf 10}, 3235 (1974).

\bibitem{soler} M. Soler, Phys. Rev. D {\bf 1}, 2766 (1970).

\bibitem{Carr1} L.H. Haddad, L.D. Carr,
  Physica D {\bf 238}, 1413 (2009).

\bibitem{Carr2} L.H. Haddad, L.D. Carr,
  EPL {\bf 94}, 56002 (2011).

\bibitem{Carr3} L.H. Haddad and L.D. Carr. New J. Phys. {\bf 17}, 113011 (2015).

 \bibitem{Carr4} L.H. Haddad, K. O'Hara, and L.D. Carr, Phys. Rev. A {\bf 91}, 043609 (2015).

 \bibitem{peleg} O. Peleg,  G. Bartal, B. Freedman, O. Manela, M. Segev,
and D.N. Christodoulides,
Phys. Rev. Lett. {\bf 98}, 103901 (2007).

\bibitem{ablo4} M. J. Ablowitz, S.D. Nixon, and Y. Zhu,
{Phys. Rev. A} {\bf 79},  053830 (2009).

\bibitem{ablo3} M. J. Ablowitz and Y. Zhu,
{Phys. Rev. A} {\bf 82},  013840 (2010).

\bibitem{PRL} J. Cuevas-Maraver, P.G. Kevrekidis, A. Saxena, A. Comech and R. Lan, Phys. Rev. Lett. {\bf 116}, 214101 (2016).

\bibitem{borism} C. Shang, Y. Zheng, B.A. Malomed, Phys. Rev. A {\bf 97}, 043602 (2018).

\bibitem{NiurkaPRE} N. R. Quintero, F. G. Mertens, F. Cooper, A. Saxena, and A. R. Bishop, Phys. Rev. E {\bf 96}, 052219 (2017).

\bibitem{dewit} B. de Wit and J. Smith, {\it Field theory in particle physics}, North Holland Physics Publishing (New York, 1986).

\bibitem{Wakano} M. Wakano, Progr. Theor. Phys. {\bf 35}, 1117 (1966).

\bibitem{2DSM1} S. M. Young and C. L. Kane, Phys. Rev. Lett. {\bf 115}, 126803 (2015).

\bibitem{2DSM2} S. V. Ramankutty et al., arXiv:1711.07165v3.

\bibitem{2DWSM} J. Ahn and B.-J. Yang, Phys. Rev. Lett. {\bf 118}, 156401 (2017).

\bibitem{ChiAP} S. Chi, Z. Li, H. Yu, G. Wang, S. Wang, H. Zhang, and J. Wang, Ann. Phys. (Berlin) {\bf 529}, 1600359 (2017).

\bibitem{Majorana} P. Zhou, J. Brand, X.-J. Liu, and H. Hu, Phys. Rev. Lett. {\bf 117}, 225302 (2016).

\bibitem{hairer} E. Hairer, P.S. N{\o}rsett, G. Wanner,
  {\it Solving ordinary differential equations I: Nonstiff
    problems}, Springer-Verlag (Berlin, 2008).

\end{thebibliography}
\end{document}